\documentclass[nocompress]{spie}  
\usepackage[]{graphicx}

\title{Design and performance of a F/\#-conversion microlens for Prime Focus Spectrograph at Subaru Telescope}

\author{Naruhisa Takato$^1$, Yoko Tanaka$^1$, James E. Gunn$^2$, Naoyuki Tamura$^3$, Atsushi Shimono$^3$, Hajime Sugai$^3$, Hiroshi Karoji$^3$, Akitoshi Ueda$^4$, Kouichi Waseda$^4$, Masahiko Kimura$^5$, and Youichi Ohyama$^5$
\skiplinehalf
$^1$Subaru Telescope, National Astronomical Observatory of Japan, 650 North A`ohoku Pl., Hilo, Hawaii 96720, USA\\
$^2$Dept. of Astrophysical Science, Princeton University, 4 Ivy Lane, Princeton, NJ 08544-1001, USA\\
$^3$Kavli Institute for the Physics and Mathematics of the Universe, University of Tokyo, 5-1-5 Kashiwanoha, Kashiwa, 277-8583, Japan\\
$^4$National Astronomical Observatory of Japan, 2-21-1 Osawa, Mitaka, Tokyo 181-8588, Japan\\
$^5$Institute of Astronomy and Astrophysics, Academia Sinica,1-4 Roosevelt Rd, Taipei 10617, Taiwan, R.O.C. 
}

\authorinfo{Further author information: (Send correspondence to N.T.)\\N.T.: E-mail: takato@naoj.org, Tel: +1 808 934 7788}
 
\begin{document} 
  \maketitle 

\begin{abstract}
The PFS is a multi-object spectrograph fed by 2394 fibers at the prime focus of Subaru telescope. Since the F/\# at the prime focus is too fast for the spectrograph, we designed a small concave-plano negative lens to be attached to the tip of each fiber that converts the telescope beam (F/2.2) to F/2.8. We optimized the lens to maximize the number of rays that can be confined inside F/2.8 while maintaining a 1.28 magnification. The microlenses are manufactured by glass molding, and an ultra-broadband AR coating ($<$1.5\% for $\lambda=0.38-1.26 \mu$m) will be applied to the front surface. 
\end{abstract}



\section{INTRODUCTION}
\label{sec:intro}  
The Prime-Focus Spectrograph(PFS) of Subaru telescope has 2394 optical fibers for recieving scientific targets at the prime foucs \cite{Sugai2014}. The PFS shares the wide-field corrector (WFC) which is designed and manufactured for the Hyper Suprime-Camera (HSC)\cite{Miyazaki2012}. Since the F-ratio of the light delivered to the prime focus is too fast for the PFS, a negative-power microlens is placed at the tip of the fiber in order to convert the F-ratio.
The F-ratio conversion is necessary for two reasons: 
(1) to increase light coupling efficiency at the input of the fiber by matching the numerical aperture (NA), 
(2) to relax the spectrograph design constraints by slowing down the F-ratio of the input beam entering the collimator.

The NA of the input beam delivered from the HSC WFCis 0.224 at the center of the field, while the NA value of the optical fiber is 0.22 with a production variation of $\pm0.01$. Also, NA is defined by the angle at which the coupling efficiency becomes 50\%.
Thus the outer part of the input light cone could be lost even at the center of the field unless we convert the F-ratio. 

The F-ratio conversion lens also makes designing the spectrograph easier.  If the F-ratio of the spectrograph collimator is F/2.2, there will be no space to accommodate two dichroic mirros necessary for our three-armed spectrograph to cover the desired wavelength range of $\lambda = 0.38 - 1.26 \mu$m \cite{Sebastien2014}. 
If we use a F-conversion lens that converts the F-ratio to f/2.8, almost all the output light from the fiber can enter a f/2.5 collimator, even after suffering focal-ratio degradation (FRD). Using a f/2.5 collimater instead of f/2.2 colimater makes it easier to design accomodating dichroics.

There are, however, several drawbacks to introducing microlenses to PFS, including (1) additional Fresnel loss, (2) possible fringing in the direction of the wavelength, and (3) increase in cost and assembly workload.

In this report, we will describe the optical and mechanical design of the microlens and some preliminary results of test-shot molding samples.

\section{DESIGN OF MICROLENS}
\label{sec:design}
\subsection{Design requirements}
\label{subsec:requiremens}
\begin{figure}[t]
   \begin{center}
   \begin{tabular}{c}
   \includegraphics[height=3cm]{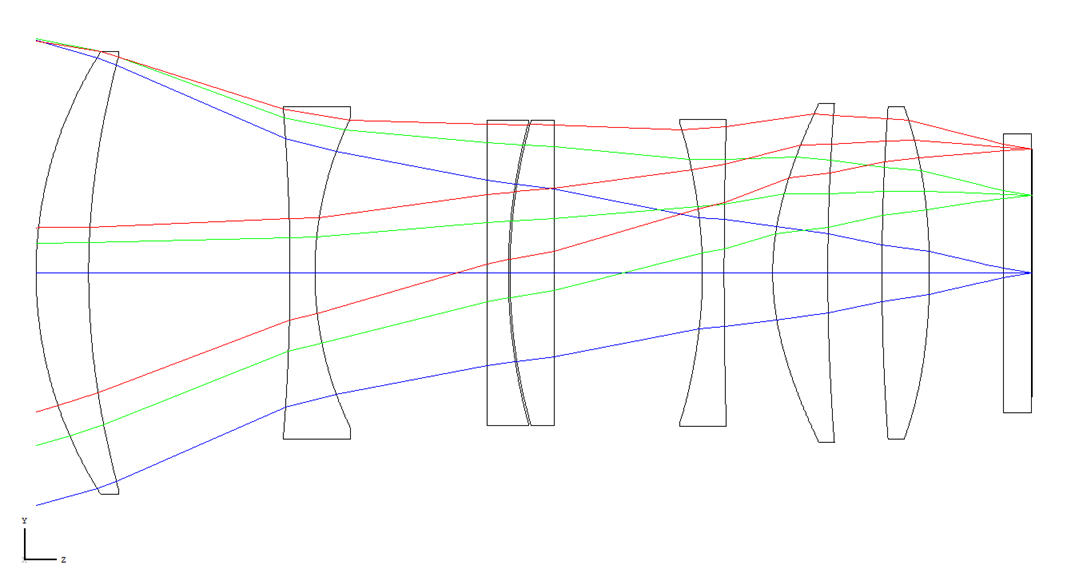}
   \end{tabular}
   \end{center}
   \caption[example] 
   { \label{fig:WFC} 
	   Wide-field corrector for Hyper Suprime-Cam and PFS of Subaru telescope. The optics is non-telecentric: incident angle of the chief ray is $6.9^\circ$ at the PFS field edge.
   }
   \end{figure}

\begin{figure}[h]
   \begin{center}
   \begin{tabular}{c}
   \includegraphics[height=12cm]{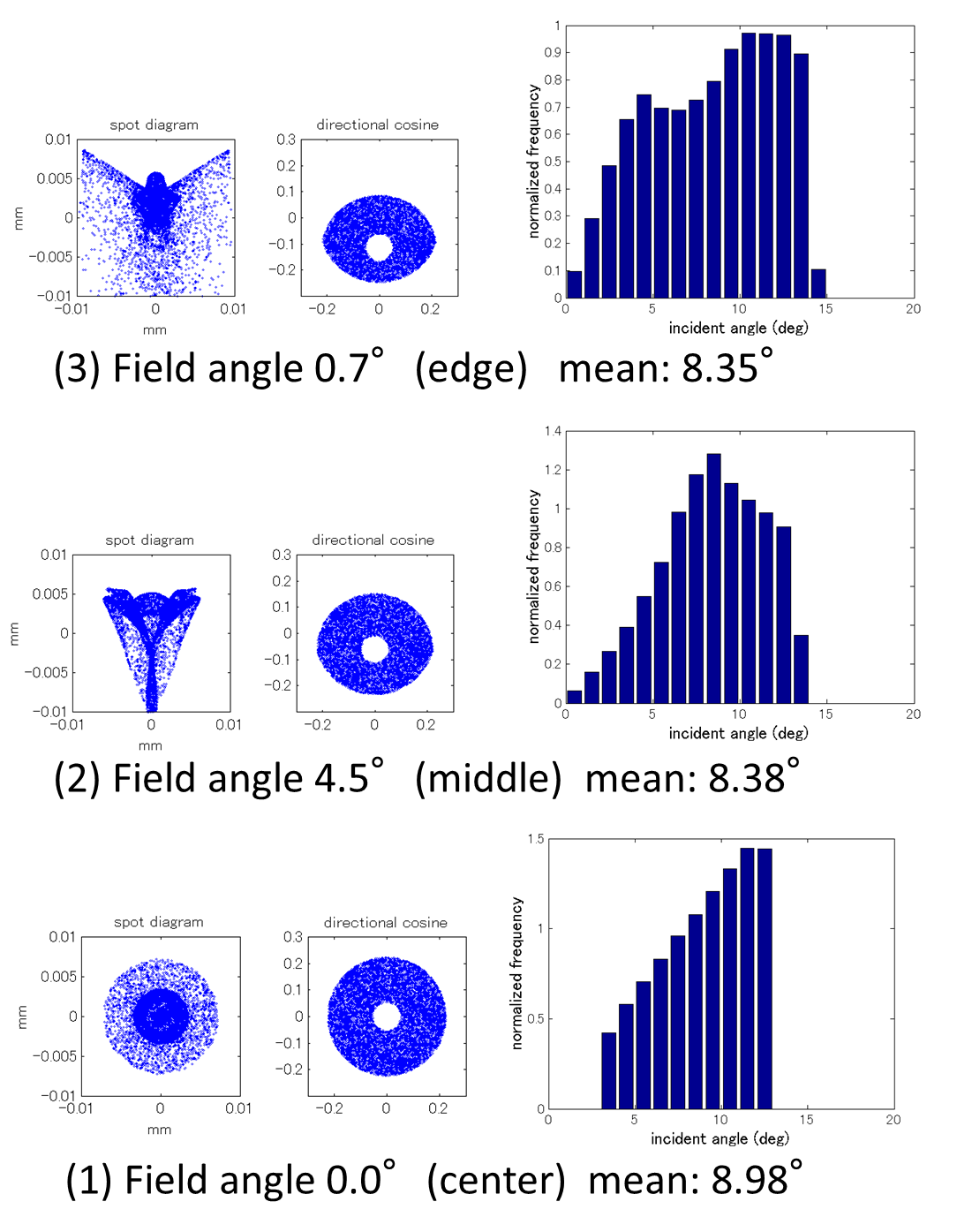}
   \end{tabular}
   \end{center}
   \caption[example] 
   { \label{fig:ray_angle} 
	Incident angle distributions at the prime focus of Subaru telescope. Note that the mean incident angle is almost the same (8-9$^\circ$) over the field of view .
   }
   \end{figure}

The focal plane of the PFS is a hexagonal region with a 1.38$^\circ$ diagonal field of view at the prime focus. The physical length of the diagonal is about 454 mm. The WFC is non-telecentric and the chief-ray of the incident beam is tilted about 6.9$^\circ$ at field edge (see Fig.~\ref{fig:WFC}). However, the mean incident angle of the rays to the focal plane is almost the same over the whole field of view because most of the rays that have large incident angles are vignetted by the lens barrel before arriving at the focal plane. The distributions of the incident angles at the center, middle and edge of the PFS field are shown in Fig.~\ref{fig:ray_angle}.


The requirements for the optical design of the microlens is below.
\begin{enumerate}
	\item Magnification should be $\times1.28$.
	\item The number of light rays $>$ F/2.8 that enter the fiber tip should be maximized.
	\item Focal plane should be on the rear surface of the microlens.
\end{enumerate}

The microlens has a magnification such that a circle with a diameter of 100 $\mu$m at the center of the WFC focal plane is projected on to the tip of the fiber as a larger circle with a diameter of 128 $\mu$m. Then, the field of view of which each fiber sees will be $1.125^{\prime\prime}$, $1.104^{\prime\prime}$, and $1.064^{\prime\prime}$ for field center, middle ($0.45^\circ$), and edge ($0.70^\circ$), respectively. This field of view is determined to maximize the signal-to-noise ratio for faint galaxies.
Requirement (2) was set because the F-ratio of the collimator of the spectrograph is F/2.5 and the fiber has FRD. 
We assumed that the nominal FRD will be $\Delta$NA$\sim0.02$. Under this assumption, the input beam at the focal plane must have NA$< 1/2\times2.5 - 0.02 \sim 0.18$, which corresponds to $\sim$ F/2.8.
The microlens is attached to a fiber tip using epoxy adhesive \cite{Santos2014} \cite{Fisher2014}.

\renewcommand{\thefootnote}{\fnsymbol{footnote}}
We chose glass-molding as the microlens fabrication method because it can minimize variations in lens dimensions\footnote[2]{This depends on the size and the shape of the lens.} The front side of the lens has a flat annular part around the concave clear aperture. This flat part will be used to determine the height of the actuator module during its assembly\cite{Fisher2014}. The manufacturer set the constraints below.
\begin{itemize}
	\item Thickness (length) of the lens should be less than 3 mm.
	\item Diameter of the lens should be larger than 1.5 mm.
	\item The lens should be a plano-concave shape.
	\item Sag of the lens should be larger than 20 $\mu$m.
	\item Molding accuracy for the lens thickness is about $\pm(10\sim20)\mu$m.
	\item The edge of the lens should have rounded chamfers of R $<$ 0.1 mm (concave side) and R $<$ 0.15 mm (flat side).
	\item There must be a transitional region (``fillet'') betweeen the clear aperture and the surrounding annular flat region.
\end{itemize}	

\subsection{Determination of the lens thickness}
\label{subsec:thickness}
\begin{figure}[]
   \begin{center}
   \begin{tabular}{c}
   \includegraphics[height=6cm]{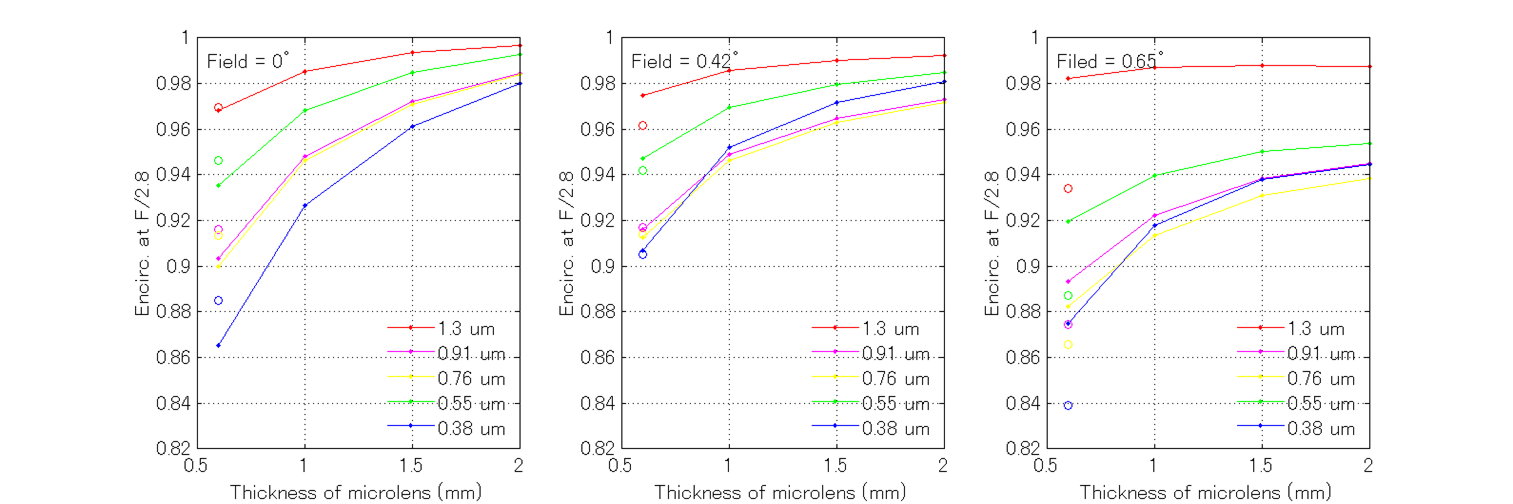}
   \end{tabular}
   \end{center}
   \caption[example] 
   { \label{fig:coupling} 
Fraction of the rays greater than F/2.8 as a function of lens thickness (open circle: optimized for the center of the fiber (aspherical), dots: plano-concave (spherical))
   }
   \end{figure}

We have investigated the performance of the microlens as  a function of lens thickness (length)  because shorter lenses are easier to mold, in general.  Fig.~\ref{fig:coupling} shows the fraction of rays coming out from the lens that is inside the F/2.8 beam cone as a function of lens length. The longer the lens, the better the performance, because the radius of the lens curvature is larger for longer lenses. Having an aspheric surface does not improve the performance of longer lenses since the aberration of the original input beam (WFC focus) is already large.  We chose the 3 mm-long microlens, which is the longest lens our manufacturer allowed, and the F-conversion performance with this lens is $>$ 97\% for most of the field and $>$ 94\% at the edge of the field.

\begin{figure}[]
   \begin{center}
   \begin{tabular}{c}
   \includegraphics[height=6cm]{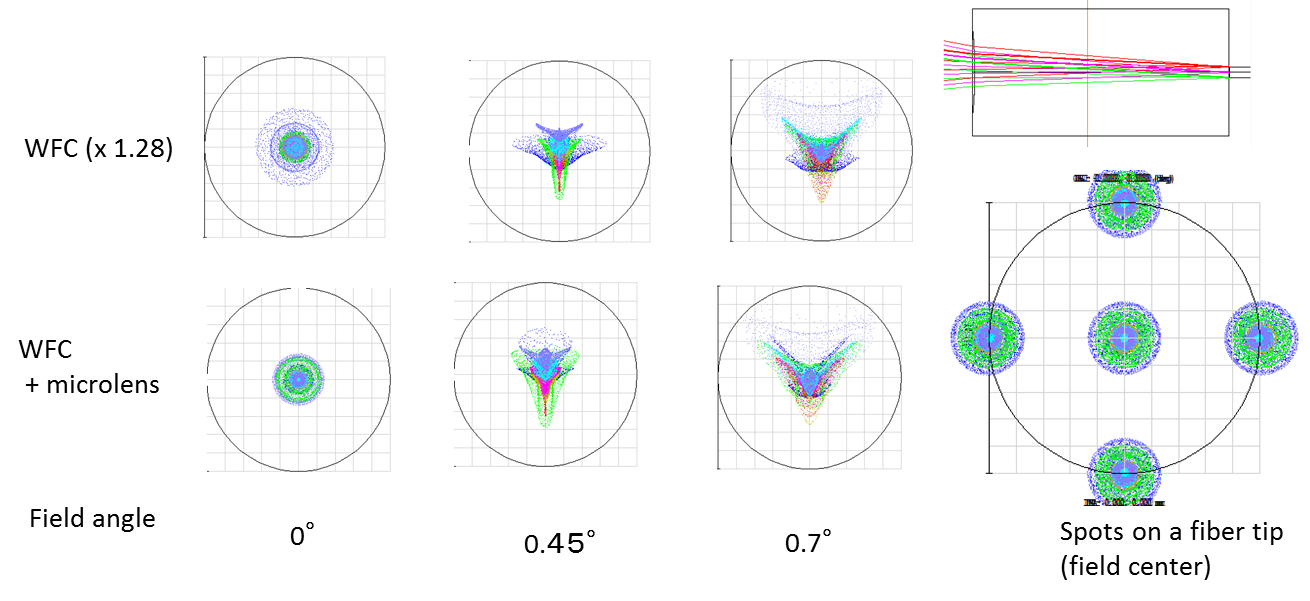}
   \end{tabular}
   \end{center}
   \caption[example] 
   { \label{fig:spot} 
Spot diagram of WFC and WFC+ microlens (circles  correspond to fiber core (128 $\mu$m)). The bottom right panel shows the spot variation at the center and the edges of the optical fiber at the WFC field center.
   }
   \end{figure}

The specifications of the microlens is shown in table~\ref{tab:spec} . Fig.~\ref{fig:spot} shows the spot diagrams of WFC with and without microlens, and its variation on the fiber tip at the center of the field of view. Almost no degradation in image quality are found.

   \begin{table}
	\label{tab:spec}
	\caption{}
	\begin{center}
	\begin{tabular}{lr}
		\hline
		Specifications & Value\\
		\hline \hline
		Center thickness (mm) & 3.000\\
		Curvature of front surface R1 (mm) & -4.7643 \\
		Position w.r.t. WFC focus (mm) & -1.3317 \\
		Clear aperture (dia.) (mm) & 1.020 \\
		Sag ($\mu$m) & 27.4 \\
		Curvature of rear surface R2 (mm) & infinity \\
		Lens diameter (mm) & 1.50 \\
		Glass material & K-VC82 \\
		Coating & Dielectric MC (R1) \\
		\hline
	\end{tabular}
	\end{center}
\end{table}

   \begin{table}
	\label{tab:spot}
	\caption{RMS spot radius ($\mu$m) with and without the microlens.}
	\begin{center}
	\begin{tabular}{lccc}
		\hline
		Field & $0^\circ$ & $0.446^\circ$ & $0.698^\circ$ \\
		\hline \hline
		WFC ($\times1.28$) & 7.74 & 10.1 & 12.9 \\
		WFC + microlens& 7.11 & 10.7 & 13.8 \\
		\hline
	\end{tabular}
	\end{center}
\end{table}

\section{TEST-SHOT SAMPLES OF GLASS-MOLD MICROLENS}
\label{sec:test_shot}
\begin{figure}[]
   \begin{center}
   \begin{tabular}{c}
   \includegraphics[height=5cm]{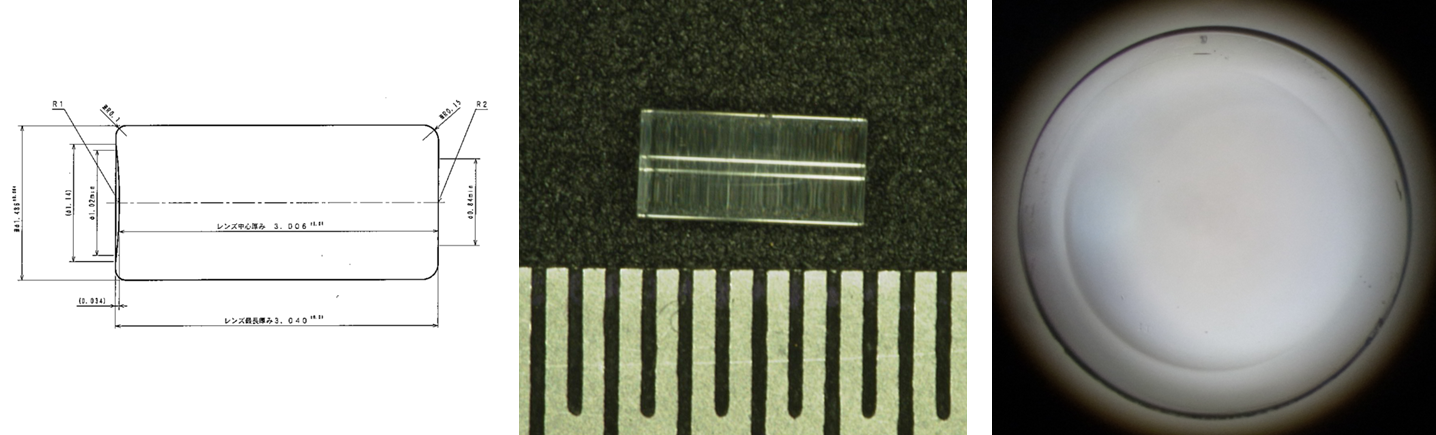}
   \end{tabular}
   \end{center}
   \caption[example] 
   { \label{fig:photo} 
	   Drawing of the microlens (left), test-shot sample of glass-mold microlens (side view) (middle), front view of the microlens (right)  Flat part surrounding the clear aperture is seen. 
   }
   \end{figure}

\begin{figure}[]
   \begin{center}
   \begin{tabular}{c}
   \includegraphics[height=6cm]{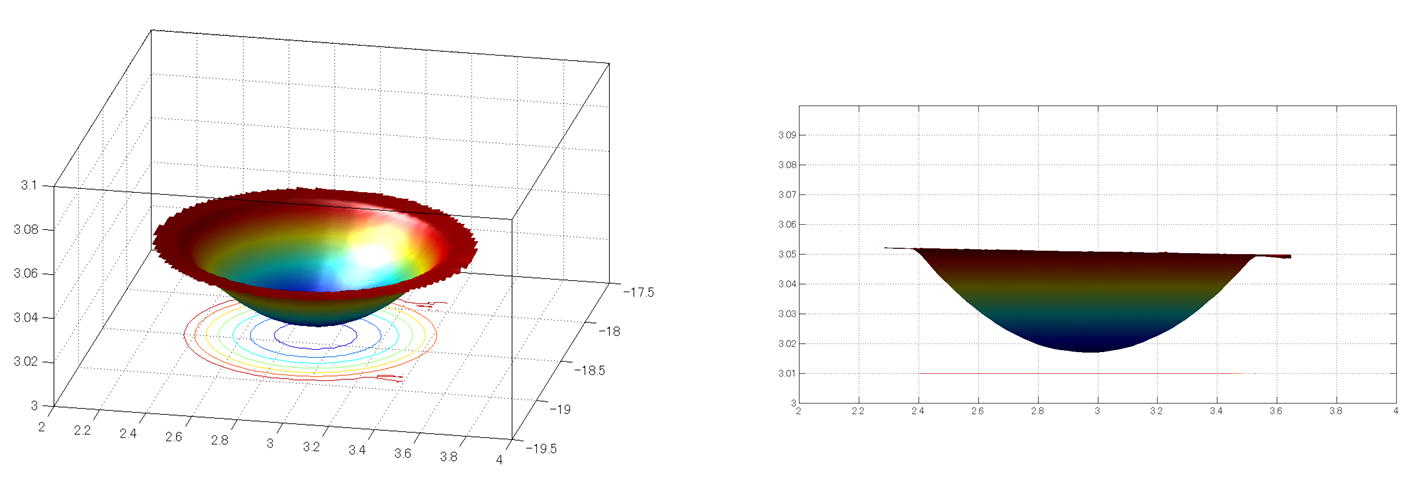}
   \end{tabular}
   \end{center}
   \caption[example] 
   { \label{fig:shape} 
	   Measured shape of the front surface of the microlens.
  }
   \end{figure}

Test-shots of the glass-mold microlens were manufactured by Panasonic Industrial Devices Nitto Co., Ltd.. Fig.~\ref{fig:photo} shows the photos of a test-shot microlens.  The front-side negative lens shape are shown in Fig.~\ref{fig:shape} which were measured using a laser probe 3D measuring instrument NH-3SP (Mitaka kohki. Co., Ltd.). The measurement points are on a 20um x 20um mesh grid. The radius of curvature is measured to be -4.740 $\pm$ 0.003 mm which is in good agreement with the specification (-4.7643 mm). The deviation from the best-fit sphere is 0.11 $\mu$m RMS and 0.57 $\mu$m PV (the PV value may be affected by dusts on the surface).

\begin{figure}[]
   \begin{center}
   \begin{tabular}{c}
   \includegraphics[height=3cm]{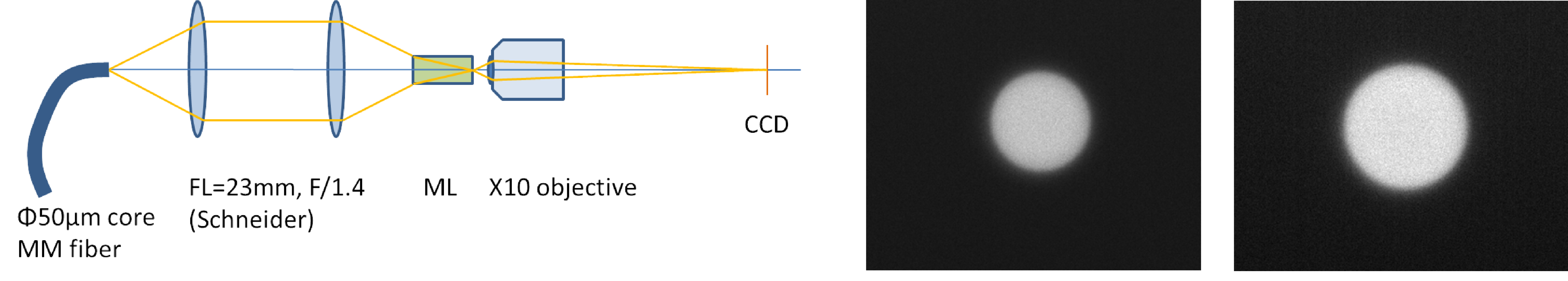}
   \end{tabular}
   \end{center}
   \caption[example] 
   { \label{fig:Mag} 
	   Setup for magnification measurements of the microlens (left), Fiber core image without (middle) and with (right) microlens. Magnification is calculated to be $\times1.29$ which is consistent with the design value of $\times1.28$. 
  }
   \end{figure}

The optical magnification is confirmed by the images through the microlens as shown in Fig.~\ref{fig:Mag}. A multi-mode fiber with $\Phi50\mu$m core diameter was illuminated at one end by a halogen lamp and the other end was used as a light source for this lens test. The tip of the fiber was re-imaged using two identical CCTV lenses whose chromatic aberrations are well corrected. The microlens that was placed at the right position with respect to the CCTV lens focus converted the F-ratio of this re-imaging beam. Then the image through the microlens was recorded by a microscope with CCD camera.
The magnification of the microlens is calculated to be $\times1.29$, which is consistent and within measurements error of the design value of $\times1.28$.


\section{COATING}
\label{sec:Coating}
\begin{figure}[]
   \begin{center}
   \begin{tabular}{c}
   \includegraphics[height=5cm]{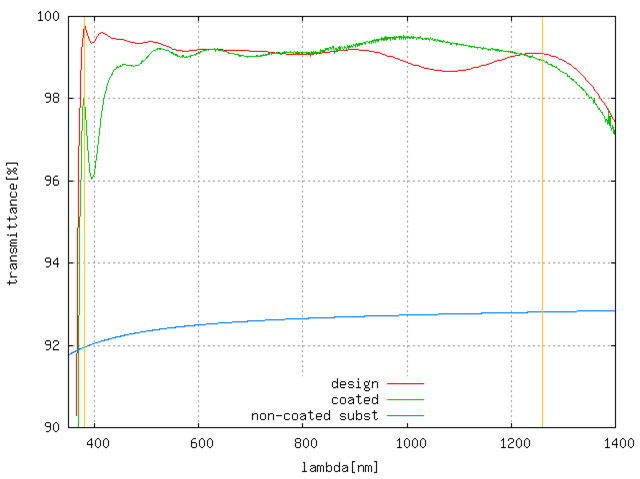}
   \end{tabular}
   \end{center}
   \caption[example] 
   { \label{fig:coating} 
	Transmittance of design and test coating on K-VC82.
  }
   \end{figure}

Since K-VC82 has a relatively high refractive index (nd=1.75550),  AR coating is necessary in order to minimize reflection loss. A multi-layer dielectric coating will be applied on the front surface of the microlens using spattering (IBS) at ATC, NAOJ. Fig.~\ref{fig:coating} shows the design and the result of a test coating on K-VC82. Reflection loss of less than 1 \% can be achieved. 

\bibliography{SPIE2012_takato}   
\bibliographystyle{spiebib}   

\end{document}